\documentclass[reprint,amsmath,amssymb, aps, pra]{revtex4-2}
\usepackage{amsmath,amssymb,physics,braket,graphicx,subfigure,epstopdf,lastpage,hyperref}
\hypersetup{colorlinks=true, urlcolor=blue, citecolor=blue, linkcolor=blue}
\allowdisplaybreaks
\begin{document}
\title{Nonclassicality versus quantum non-Gaussianity of photon subtracted displaced Fock state}
	\author{Deepak}
	\email{deepak20dalal19@gmail.com}
	\affiliation{Department of Mathematics, J. C. Bose University of Science and Technology,\\ YMCA, Faridabad 121006, India}
	\author{Arpita Chatterjee}
	\email{arpita.sps@gmail.com}
	\affiliation{Department of Mathematics, J. C. Bose University of Science and Technology,\\ YMCA, Faridabad 121006, India}
	\date{\today}
	\begin{abstract}

In this paper, a quantitative investigation of the non-classical and quantum non-Gaussian characters of the photon-subtracted displaced Fock state $\ket{\psi}=a^kD(\alpha)\ket{n}$, where $k$ is the number of photons subtracted, $n$ is Fock parameter, is performed by using a collection of measures like Wigner logarithmic negativity, linear entropy potential, skew information based measure, and relative entropy of quantum non-Gaussianity. It is noticed that the number of photons subtracted ($k$) changes the nonclassicality and quantum non-Gaussianity in a significant amount in the regime of small values of the displacement parameter whereas the Fock parameter ($n$) presents a notable change in the large regime of the displacement parameter. In this respect, the role of the Fock parameter is found to be stronger as compared to the photon subtraction number. Finally, the Wigner function dynamics considering the effects of photon loss channel is used to show that the Wigner negativity can only be {detected} by highly efficient detectors.
\end{abstract}
%\begin{keywords}
% {nonclassicality, quantum non-Gaussianity, displaced Fock state, photon subtraction, Wigner function}
%\end{keywords}
\maketitle
	\section{Introduction}
\label{intr}

{Quantification} of quantum non-Gaussianity and nonclassicality of a radiation field has been considered extensively in recent days. This attention is justified as with the appearance of quantum information science, it is appreciated that nonclassicality and quantum non-Gaussianity are two main features of the quantum world, which can direct to quantum supremacy \cite{png1}. A theoretical work observed that the photon-subtracted Gaussian state is non-classical if and only if the initial Gaussian state is non-classical, whereas the scenario is different for the photon-added, photon-added-then-subtracted, and photon-subtracted-then-added Gaussian states, they are always nonclassical independent of the initial Gaussian state \cite{pyyf9}. This investigation motivates us to study the non-classical as well as quantum non-Gaussian characteristics of the photon-subtracted displaced Fock state.

Quantification of nonclassicality and quantum non-Gaussianity can be approached in different ways. In any case, no particular measure of nonclassicality or quantum non-Gaussianity can be constructed. Here we consider a measure that can estimate the quantum non-Gaussianity of a field state which is an indispensable resource for quantum information processing. Genoni et. al. \cite{genoniparis} addressed two entanglement-distillation
protocols of quantum non-Gaussianity depending on the Hilbert-Schmidt distance and the quantum relative entropy (QRE)
between the state undergoing the experiment and a reference Gaussian state. They found that the quantum non-Gaussianity appears to be dependent on the protocol itself, and in either of the two distillation protocols, the amount of the gained entanglement at each
step is monotonous with the quantum non-Gaussianity of the initial low-entangled state. They also illustrated that in the bipartite setting, there
is a connection between correlations and quantum non-Gaussianity, that is, at fixed covariance matrix quantum non-Gaussian
states have more correlations and this excess of correlations is
proportional to the amount of quantum non-Gaussianity that the quantum state loses during partial trace operation or decoupling. These results on the robustness of quantum non-Gaussian
entanglement in noisy Markovian channels suggest that there are regimes where quantum non-Gaussian resources can be exploited to improve quantum communication protocols. The findings of \cite{genoniparis} pave the way for further research and indicate that a detailed understanding of the geometrical and analytical structures underlying the quantum non-Gaussian features of states and operations could be an extremely useful tool for the successful implementation of continuous-variable (CV) quantum information processing.

Since the advancement of continuous-variable quantum technology, most of the protocols designed for finite-dimensional Hilbert spaces have been first
implemented in the CV setting by using Gaussian states for a number of reasons. Gaussian states are experimentally produced with a high degree of control,
especially in quantum optics, and Gaussian measurements may be effectively implemented in different situations. Moreover, Gaussian states, being a member of an infinite-dimensional Hilbert space, are easy to handle from a theoretical point of view as these states are fully described by the first and second moments of
the canonical operators \cite{fagp1,fagp2,fagp3,fagp4,fagp5,fagp6}. However, it has recently come to light that there are situations where quantum non-Gaussianity, in the form of quantum non-Gaussian states or quantum non-Gaussian operations, is required to complete some relevant tasks in quantum information processing. For example, quantum non-Gaussianity is essential for realizing entanglement distillation \cite{gp9,gp10,gp11}, quantum error correction \cite{gp12}, and cluster-state quantum computation \cite{gp13,gp14}. Additionally, quantum non-Gaussian measurements and/or quantum non-Gaussian states are crucial for detecting violations in CV loophole-free Bell tests \cite{gp15,gp16,gp17,gp18,gp19,gp20,gp21,gp22,gp23,gp24}. Also, quantum non-Gaussian states or quantum non-Gaussian operations can be used to improve quantum teleportation \cite{gp25,gp26,gp27} and quantum cloning of coherent states, respectively.

Any quantum state $\rho$ can be described in respect of the well-known Glauber-Sudarshan $P$ function as \cite{png2,png3}
\begin{equation}
\label{pal}
\rho = \int{d(P(\alpha))}\ket{\alpha}\bra{\alpha}.
\end{equation}

If $P(\alpha)$ skips to be a probability distribution function, the state is non-classical. In other words, a state that cannot be precisely written in terms of a statistical mixture of coherent states is known as a non-classical state. Recently, the significance of non-classical states has been improved considerably as in the dominion of meteorology and
information processing \cite{png4}, enhancing the performance of the devices is of high demand and it is found that the desired enhancement is possible only using quantum resources. Non-classical states are very important for acquiring quantum advantage and thus, they are considered as the basic building blocks for quantum-enhanced devices having advantages over their corresponding classical variants. It can be mentioned that the non-classical states not only detect gravitational waves in LIGO experiment \cite{png5}, they are found to be essential for quantum teleportation \cite{png6,png7,png8}, quantum key distribution \cite{png9,png10,png11}, quantum computation \cite{png12,png13} and so on. As nonclassicality directs to quantum advantage, it is time to quantify the amount of nonclassicality and the quantum advantages it can provide. In 1987, Hillery presented a distance-based measure \cite{png14}, however, there were numerous computational challenges related to that. Following that, Mari et. al. \cite{png15} introduced a measure of nonclassicality in terms of the operational distinguishability of a quantum state with a positive Wigner function, given by
$\displaystyle \eta(\rho) =\min_{\omega\in \mathcal{C} } ||\rho-\omega||$ where $\mathcal{C}$ denotes the set of all quantum states with positive Wigner function and $||. ||_1$ is the trace norm. In 1991, Lee proposed a measure for the quantification of nonclassicality known as nonclassical depth \cite{png16} (for a brief review see \cite{png17}). These reviews unveiled that quantum non-Gaussianity inducing operations, e.g., photon addition, photon subtraction, and their combinations \cite{png18,png19} can introduce and/or increase the amount of nonclassicality in an arbitrary quantum state.

Filip et. al. \cite{filip2011} proposed a novel criterion for uncovering quantum states of a harmonic oscillator with a positive Wigner function which could not be expressed as a convex mixture of Gaussian states. Quantum non-Gaussian states can be defined as the states which cannot be defined in terms of a probabilistic mixture of Gaussian states \cite{genoni,kuhn}. In a recent year, Chabaud et. al. \cite{chabaud} introduced the so-called stellar formalism that characterizes non-Gaussian quantum states by the distribution of the zeros of their Husimi $Q$ function in phase space. They have studied the topological properties of the stellar hierarchy with respect to the trace norm in detail. In addition, all the Gaussian states can be defined by the first- and second-order moments, which means that the mean and covariance matrix of the Gaussian states provide their complete information. We can define
a complex hull $\mathcal{G}$ with the classical probability distribution $P_{cl}(\lambda)$
\begin{equation}
\label{pall}
\rho = \int{d(P_{cl}(\lambda))}\ket{\psi_G(\lambda)}\bra{\psi_G(\lambda)}.
\end{equation}
in the Hilbert space $\mathcal{H}$. This set includes all the Gaussian states and some quantum non-Gaussian states. Interestingly, the quantum non-Gaussian states obtained as the statistical mixtures of Gaussian states in the form \eqref{pall} have minimal applications due to their origin in classical
noise \cite{genoni,kuhn}. On the other hand, quantum non-Gaussian states $\rho$ in $\mathcal{H}$ are the states which do not belong to the complex hull $\rho \notin \mathcal{G}$. It is worth studying the quantification of quantum non-Gaussianity in such a state as these states can be used as more robust
resources compared to the Gaussian states (\cite{meni,adesso,al} and references therein). Specifically, no-go theorems are limiting
the applications of Gaussian operations and Gaussian states in entanglement distillation, quantum error correction, quantum computing, and quantum bit commitment. The use of quantum non-Gaussian operations is known to provide advantages in quantum computation, quantum communication, quantum metrology, etc. Although, in what follows, these exciting features of nonclassicality and quantum non-Gaussianity instigated us to study how the nonclassicality and quantum non-Gaussianity change with different state parameters of photon subtracted displaced Fock state (PSDFS).

\begin{figure}[htb]
\centering
\includegraphics[scale=.4]{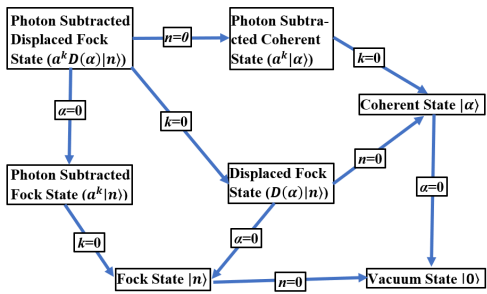}
\caption{(color online) Some special cases of photon subtracted displaced Fock state.}
\label{figpsdfs}
\end{figure}

There are several reasons for choosing PSDFS as the state of interest to study the measures of nonclassicality and quantum non-Gaussianity. Firstly, in a different limiting context, this state provides a set of well-known quantum states having many applications. The PSDFS is the well-known state obtained by subtracting photons from the displaced Fock state (DFS). Different features of DFS have been studied extensively \cite{oliveira1,oliviera2,chatterjee1} in literature. Further, it is observed that quantum non-Gaussianity-inducing operators significantly affect the signatures of nonclassicality of DFS viewed through different witnesses \cite{alam1}. Therefore, it is appropriate to consider the PSDFS for studying the nonclassicality and quantum non-Gaussianity features, using Wigner negativity, linear entropy potential, skew information-based measure, Wigner logarithmic negativity, and relative entropy measure of quantum non-Gaussianity. Various limiting cases of PSDFS are described in Fig.~\ref{figpsdfs}. PSDFS reduces to displaced Fock state if $k=0$ and photon subtracted coherent state if $n=0$. Further, displaced Fock state can be converted to coherent and Fock state by taking $n=0$ and $\alpha=0$, respectively. Again photon subtracted coherent state reduces to the coherent state by taking $k=0$ and hence the final coherent and Fock states are further reduced to a vacuum state $\ket{0}$ by putting $\alpha=0$ and $n=0$, respectively.

\begin{figure}[htb]
\centering
\includegraphics[scale=.3]{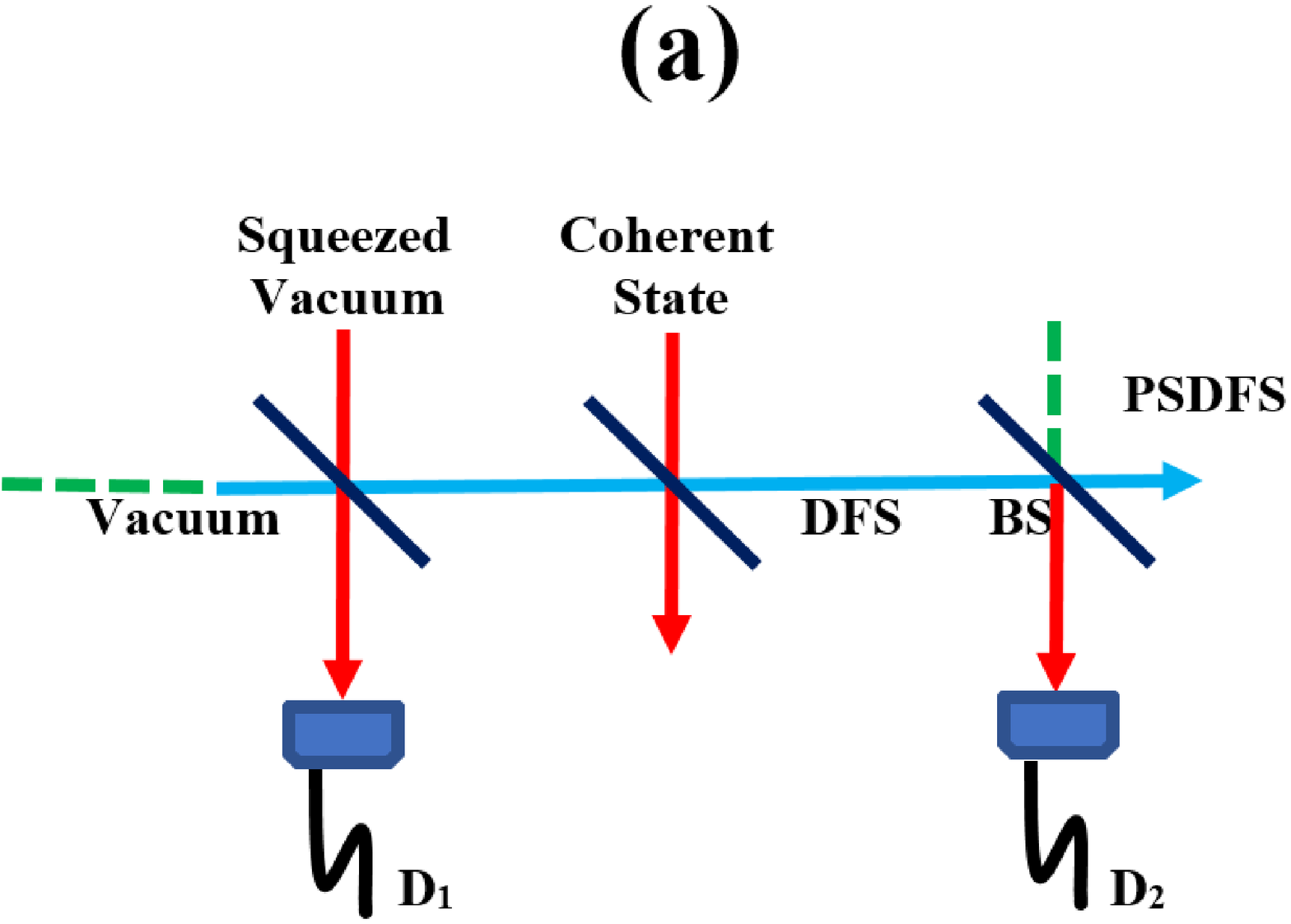}
\includegraphics[scale=.3]{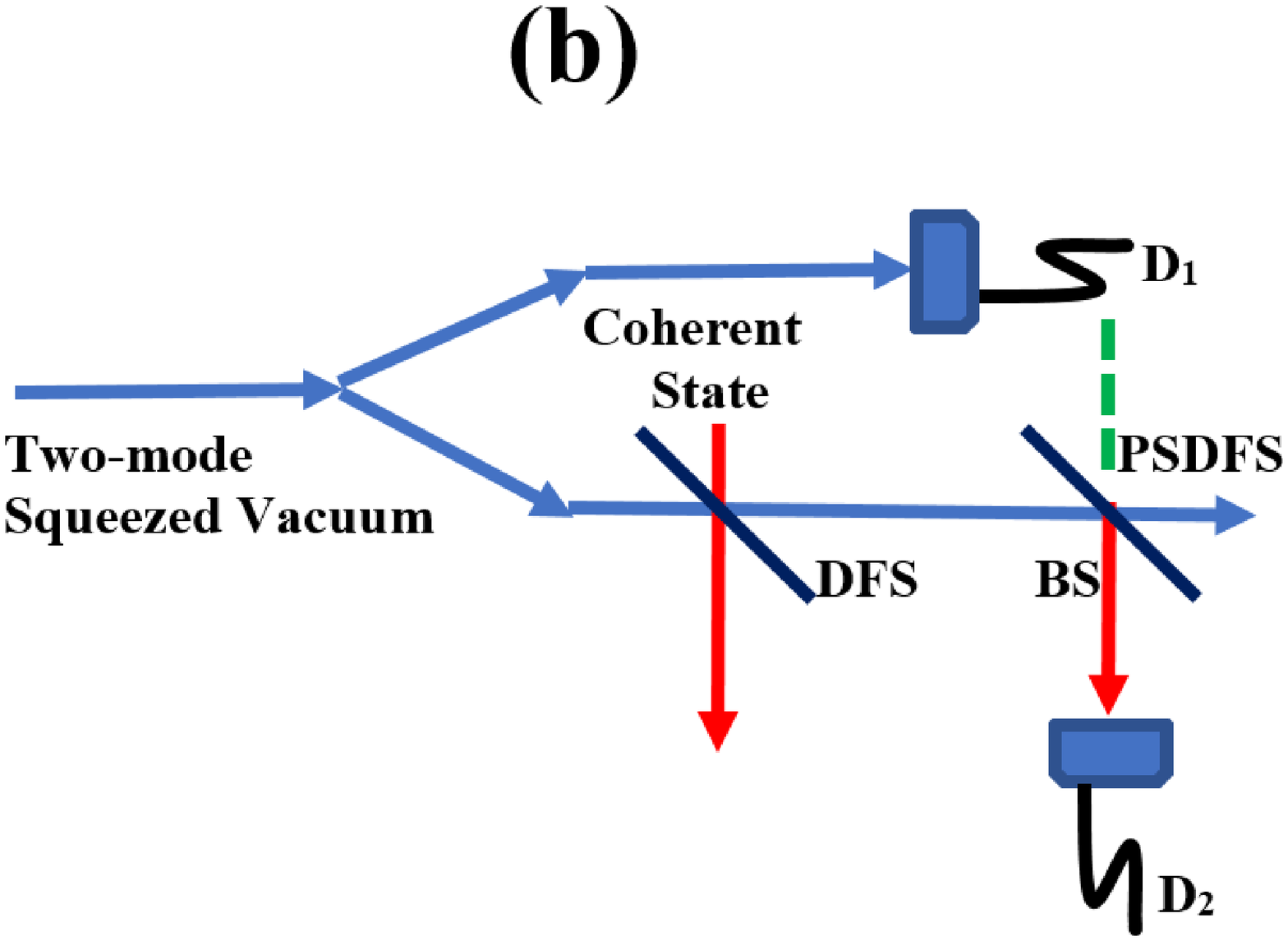}
\caption{(color online) Generation of PSDFS using different quantum states and beam-splitters.}
\label{psdfsgen}
\end{figure}

It is worth noting that DFS has been generated experimentally by superposing a Fock state with a coherent state
on a beam-splitter \cite{lvosky}. In this way, a schematic diagram for generating the PADFS and PSDFS using single-mode and two-mode squeezed vacuum states was proposed in \cite{png19}. Using three (two) highly transmitting beam-splitters, a conditional measurement of single photons at both detectors $D_1$ and $D_2$ as in Fig.~\ref{psdfsgen}{\color{blue}(a)} (Fig.~\ref{psdfsgen}{\color{blue}(b)})
would result in a single photon subtracted DFS as output from a single-mode (two-mode) squeezed vacuum state. Repeating the process $k$ times, $k$-PSDFS can be built in the lab.

The structure of the paper is as follows. In section \ref{qsi}, we find the analytical expression for PSDFS, and using it we further calculate the Wigner function. In section \ref{ncngc}, we discuss different measures of nonclassicality and quantum non-Gaussianity of PSDFS and perform a comparison. In section \ref{loss}, the dynamics of the Wigner function for PSDFS over the photon loss channel is described. The paper ends with a conclusion in the section \ref{conclusion}.

\section{Photon subtracted displaced Fock state and its Wigner function}
\label{qsi}

A beam splitter whose second port is fed by a highly excited coherent state can effectively approximate the displacement operator $D(\alpha)$ on any quantum state of the radiation field \cite{mgaparis}. Thus, a displaced Fock state (DFS) is analytically defined as $\ket{\alpha, n}=D(\alpha)\ket{n}$, where $D(\alpha)$ is the displacement operator and $\ket{n}$ is the Fock
state with $n$ photons. This nonclassical state, while $D(\alpha)$ operating on vacuum state $\ket{0}$ provides a classical coherent state $\ket{\alpha}$. In the present work, we deal with photon-subtracted displaced Fock state (PSDFS) obtained by applying the quantum non-Gaussianity introducing annihilation operator $k$-times on DFS. We attempt here to quantify the amount of nonclassicality present in the PSDFS and then discuss its adherence with quantum non-Gaussianity in detail. The $k$ photon subtracted displaced Fock state can be written as follows (see Appendix \ref{appdspsdfs})
\begin{eqnarray}\nonumber
\label{ieq}
\ket\psi & = & N a^kD(\alpha)\ket{n}\\
& = & \sum_{m=0}^\infty C_m(n,\alpha,k)\ket{m}
\end{eqnarray}
where
\begin{eqnarray*}
C_m(n,\alpha,k) = Ne^{-\frac{|\alpha|^2}{2}}\alpha^{m-n+k}\sqrt{\frac{n!}{m!}}L_n^{m+k-n}(|\alpha|^2)
\end{eqnarray*}
with normalization factor $N$ as
\begin{eqnarray*}
N=\left[\sum_{r=0}^k\binom{k}{r}^2\frac{n!}{(n-r)!}|\alpha|^{2(k-r)}\right]^{-1/2}
\end{eqnarray*}
and $L_n^m(x)$ is the associated Laguerre polynomial.

The Wigner function for an arbitrary quantum state with density matrix $\rho$ can be described as \cite{wigner}
\begin{eqnarray}
W(\gamma,\gamma^*)& =& \frac{2}{\pi^2} e^{-2|\gamma|^2}\nonumber\\&&\times\int{d^2\lambda\,\langle-\lambda|\rho|\lambda\rangle\,\exp[-2(\gamma^*\lambda-\gamma\lambda^*)]}\nonumber\\
\end{eqnarray}
Using this representation, the Wigner function of PSDFS is calculated
\begin{align}
\label{swf}
W(\gamma,\gamma^*) &= \frac{2N^2\exp(-2|\eta|^2)}{\pi}\sum_{p,q=0}^k\binom{k}{p}\binom{k}{q}\alpha^{*k-p}\alpha^{k-q} n!\nonumber\\&\times\sum_{r=0}^{n-p}\frac{(-2)^{p-q+r}\eta^{*p-q}}{r!(p-q)!(n-p-r)!} ~_1F_1(-r;p-q+1;2|\eta|^2),
\end{align}
where $\eta=\alpha-\gamma$, $_1F_1(a;b;x)$ is the generalized hypergeometric function (see Appendix \ref{awf} for details).

\begin{figure*}[htb]
\centering
\includegraphics[scale=1.25]{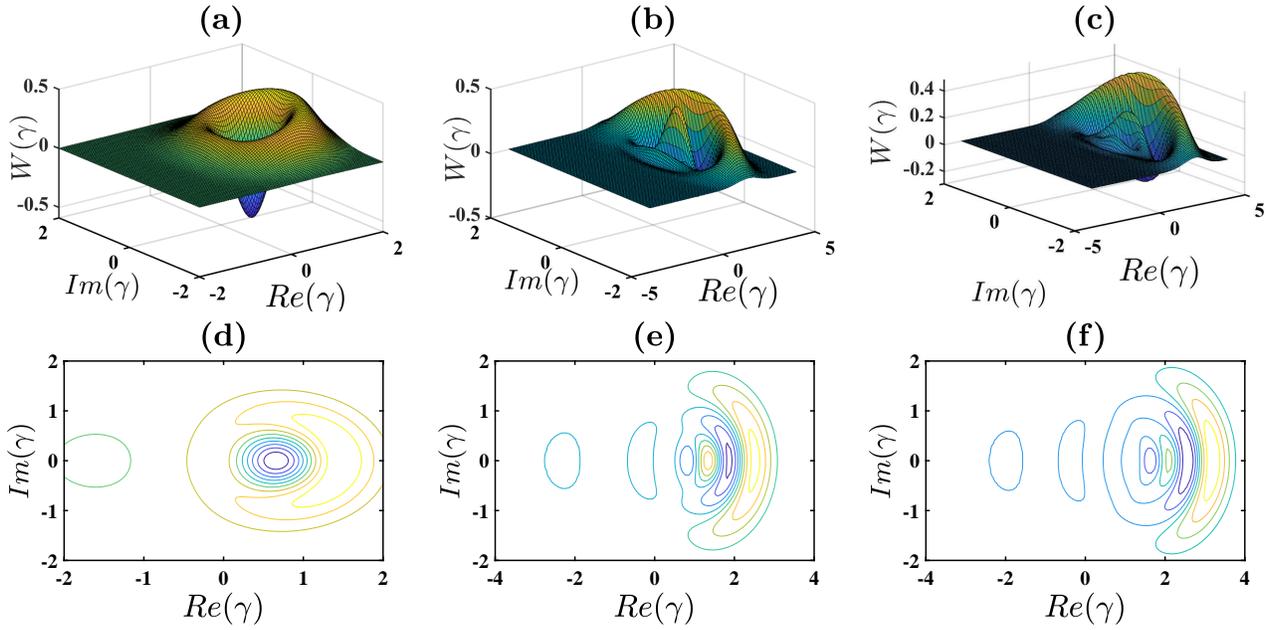}
\caption{(Color online) Variation of Wigner function with regard to real and imaginary $\gamma$ and for (a) $k=1,\,n=3,\,\alpha=0.5$, (b) $k=2,\,n=4,\,\alpha=1$, (c) $k=3,\,n=5,\,\alpha=1.5$, (d)-(f) are corresponding contour plots.}
\label{figwf}
\end{figure*}

The Wigner function is plotted here for different state parameter values. The existence of non-classical and quantum non-Gaussian features in the photon-subtracted displaced Fock state can be studied in terms of the Wigner function. The negative region of the Wigner function indicates the existence of nonclassicality. In addition to that, Hudson's theorem proved \cite{hud} that any pure quantum state having a non-negative Wigner function is necessarily Gaussian. The quantum state in which we have complete knowledge of the quantum system is known as a pure state. Different distributions of pure states can produce equivalent mixed states, and the mixed state is the combination of probabilities of the information about the quantum state \cite{pams}. PSDFS are by definition pure states. This infers that the negative values of the Wigner function witness the quantum non-Gaussianity in the state. Moreover, the ``quantum non-Gaussianity criteria" \cite{genoni} derived from the Hudson's theorem, sets out a lower bound on the Wigner function at the origin of phase space for a pure Gaussian state. It certifies that for any given pure single-mode Gaussian state $\ket{\psi_G}$, the value of the Wigner function at the origin of the phase space is bounded from below by $\frac{2}{\pi}\exp\{-2n(1+n)\}$ where $n=\bra{\psi_G}a^\dag a\ket{\psi_G}$. Here, the Wigner function \eqref{swf} satisfies the lower bound condition for the parametric values taken in Fig.~\ref{figwf}.

It may be noted from \eqref{swf} that the Wigner function is of Gaussian form for $k=n=0$ as it corresponds to the Wigner function of the coherent state. It is clear from Fig.~\ref{figwf} that the non-zero value of the displacement parameter can influence the quantum non-Gaussian behavior of the state if either $k\neq 0$ or $n\neq 0$. Moreover, the non-zero Fock parameter and photon subtraction lead to the quantum non-Gaussianity of the Wigner function having a Gaussian factor. The negative region of the Wigner function in Fig.~\ref{figwf} clearly exhibits the non-classical as well as the quantum non-Gaussian features of the PSDFS. It is also observed that with an increase in values of $k$, $n$, and $\alpha$, the negative values of the Wigner function decrease.

\subsection{Reconstruction algorithm for Wigner function} To view the applicability of the results in a practical experiment, we have considered an optical state reconstruction algorithm using the Wigner function description \cite{tomo}. In the
case of homodyne tomography, it is convenient to represent the reconstructed state in the form of the phase-space quasiprobability density, the Wigner function as
\begin{eqnarray}
\label{tomonew}
& & W_{\rho}(Q, P)\\\nonumber & = & \frac{1}{2\pi}\int_{-\infty}^{\infty}\langle Q+\frac{1}{2} Q'|\rho|Q-\frac{1}{2} Q'\rangle e^{-i P Q'} dQ'
\end{eqnarray}
The experimentally measured histogram $Pr(Q_\theta, \theta)$ is the integral
projection of the Wigner function onto a vertical plane
oriented at angle $\theta$ to the $Q$ axis as
\begin{eqnarray}
& & Pr(Q_\theta, \theta)\\\nonumber & = & \int_{-\infty}^{\infty} W_\textrm{det} (Q_\theta \cos\theta-P_\theta \sin\theta, Q_\theta \sin\theta+P_\theta \cos\theta) dP_{\theta}
\end{eqnarray}
The ``detected" Wigner function $W_\textrm{det}$ corresponds to the ideal Wigner function \eqref{tomonew} for a loss-free detector, and for a detector
with quantum efficiency $\eta$, it is obtained from the latter via a convolution as
\begin{eqnarray}\nonumber
& & W_\textrm{det}(Q, P)\\ & = & \frac{1}{\pi(1-\eta)}\int_{-\infty}^{\infty} \int_{-\infty}^{\infty}W(Q', P')\\\nonumber
& & \times\exp\Big[-\frac{(Q-Q'\sqrt{\eta})^2+(P-P'\sqrt{\eta})^2}{1-\eta}\Big] dQ' dP'
\end{eqnarray}

\begin{figure*}[htb]
\centering
\includegraphics[scale=1.25]{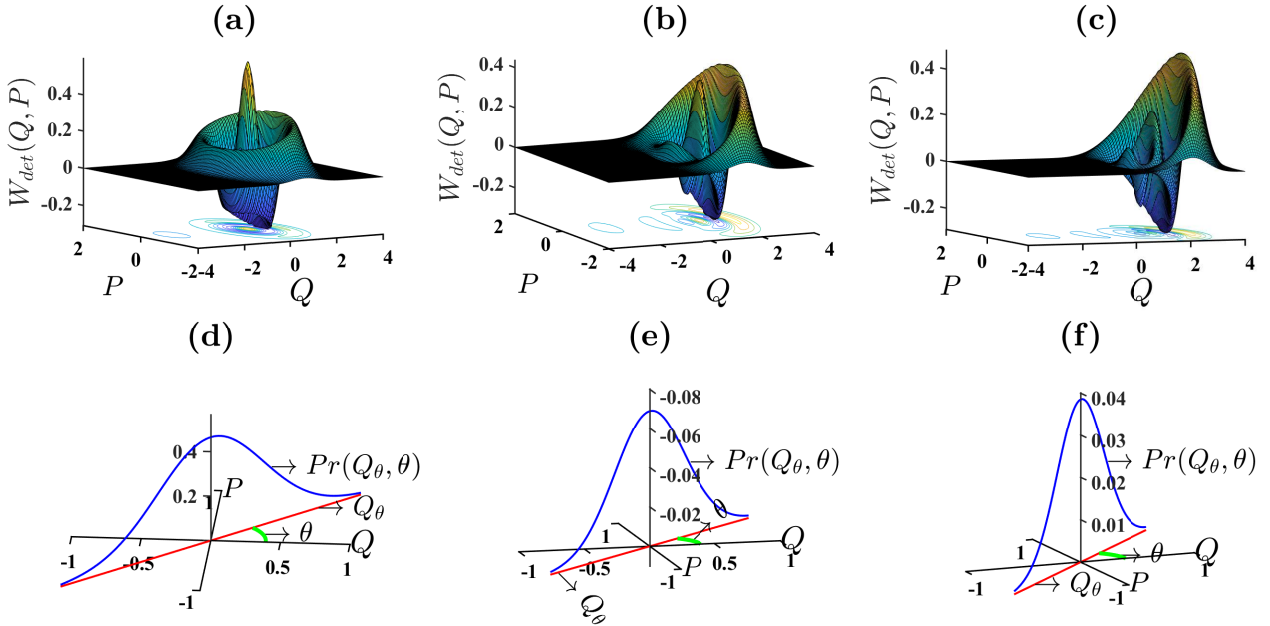}
\caption{(Color online) Variation of reconstructed Wigner function $W_\textrm{det}(Q,P)$ with respect to $Q$ and $P$ with $\eta=0.5,\, \theta=\pi/4$ and (a) $k=1,\,n=3,\,\alpha=0.5$, (b) $k=2,\,n=4,\,\alpha=1$, (c) $k=3,\,n=5,\,\alpha=1.5$, (d)-(f) are the corresponding projection curve $Pr(Q_\theta, \theta)$ and quadrature line $Q_\theta$. The experimentally measured field quadrature probability density $Pr(Q_\theta, \theta)$ is
the integral projection of the ideal Wigner function $W(Q, P)$ onto a vertical plane defined by the phase of the local oscillator.}
\label{wpff}
\end{figure*}
{In Figs.~\ref{wpff}(a)-(c), we have plotted the detected Wigner function $W_\textrm{det}(Q,P)$ with the same parametric values as in Fig.~\ref{figwf}. In a practical experiment, the photodiodes in the homodyne detector are not $100\%$ efficient, i.e., they do not transform every incident photon into a photoelectron. This leads to a distortion of the quadrature noise behavior which needs to be compensated for in the reconstructed state. The reconstructed Wigner function corresponds to the detectors with a nonunitary quantum efficiency $\eta$ (= 0.5). It is clear from Fig.~\ref{wpff} that the Wigner function restored using tomography is adequate to confirm the nonclassical and quantum non-Gaussian features of the presented optical state. The field quadrature probability density $Pr(Q_\theta, \theta)$   observing the field quadrature equal to $Q_\theta$ is described in Figs.~\ref{wpff}(d)-(f). This figure signifies the error while the Wigner function of the photon subtracted displaced Fock state is reconstructed by performing a full quantum state tomography. The integral projection of the ideal Wigner function $Pr(Q_\theta, \theta)$ quantifies the error when the results are applied to an actual physical system.}

\section{Measures of nonclassicality and quantum non-Gaussianity}
\label{ncngc}

Photon subtraction is an extensively used operation to design quantum states and is employed repeatedly to transform a classical and Gaussian state to a non-classical and quantum non-Gaussian one. Consequently, this operation can be treated as a quantum non-Gaussianity injecting operator. It can also be used for introducing nonclassicality into a quantum state by hole burning process \cite{png44,png45}. In this section, we find out the variation of nonclassicality and quantum non-Gaussianity present in the PSDFS while different state parameters (number of photons subtracted ($k$), displacement parameter ($\alpha$), Fock parameter ($n$)) are changed. The current study is of utmost importance as it may help to identify the appropriate state parameter values for a quantum-designed state which is to be used for executing a quantum calculation, correspondence, or metrology task that requires non-classical and additionally quantum non-Gaussian states. For the measurement of nonclassicality \cite{png}, the closed-form analytic expression of an entanglement potential (referred to as linear entropy potential), skew information-based measure, and Wigner logarithmic negativity are obtained in this section. Further, Wigner logarithmic negativity and relative entropy of quantum non-Gaussianity are calculated to estimate quantum non-Gaussianity. In the following subsections, we briefly present these measures and then explicitly describe how the amount of nonclassicality and quantum non-Gaussianity in PSDFS quantified by these conditions change with the state parameters.

\subsection{Linear entropy potential}

Asboth \cite{pmng46} proposed a new measure of nonclassicality based on the fact that if a single-mode non-classical (classical) state is injected into one input port of a beam-splitter (BS) and a vacuum state is embedded from the other end, the output two-mode state should be entangled (separable). Therefore, the nonclassicality of the input single-mode state (other than the vacuum state inserted in the BS), either classical or non-classical, can be indirectly measured in terms of an entanglement measure. A variety of quantitative entanglement measures are available which can be used to estimate the nonclassicality of the input single-mode state. When an entanglement measure following Asboth's approach is used to assess the single-mode nonclassicality, it is referred as entanglement potential in correspondence with the Asboth's terminology. In particular, if concurrence (linear entropy) is used to quantify the nonclassicality of the single-mode input state by measuring the entanglement of the two-mode output state leaving from the BS, the nonclassicality criteria is named as concurrence potential (linear entropy potential) \cite{pmng17}.

The linear entropy for a bipartite state $\rho_{AB}$ is defined with respect to in terms of the reduced subsystem as
\begin{equation}
\label{ile}
L_E = 1-\text{Tr}(\rho_B^2)
\end{equation}
where $\rho_B$ is the partial trace of $\rho_{AB}$ over the subsystem $A$. Therefore $L_E$ corresponds to 1 (0) for a maximally entangled (separable) state, and a non-zero value for an entangled state, in general, \cite{png47,png48}. To compute the linear entropy potential, we require the post-beam-splitter state $\rho_{AB}$ which is originated if PSDFS and vacuum states are mixed at a beam-splitter. \\
The two-mode state exiting from the beam-splitter when the Fock state is injected at one port and vacuum at the other can be expressed as follows
\begin{equation}
\ket{n}\otimes\ket0=\ket{n,0}\,\xrightarrow{BS}\,\frac{1}{2^{n/2}}\sum_{j=0}^n \sqrt{\binom{n}{j}}\ket{j,n-j}
\end{equation}
This equation can be used to find the density matrix $\rho_{AB}$ as follows
\begin{align}
\ket\phi_{AB}& = \sum_{m=k}^\infty C_m(n,\alpha,k)\,\left(\frac{1}{\sqrt{2}}\right)^{m}\,\\
&\times\sum_{j=0}^{m-k}\sqrt{\binom{m-k}{j}}\ket{j,m-k-j}\nonumber
\end{align}
and then
\begin{align}\nonumber
\rho_{AB}\\ & = \sum_{m,m'=k}^\infty C_m(n,\alpha,k)C_{m'}^*(n,\alpha,k)\left(\frac{1}{\sqrt{2}}\right)^{m+m'} \\ & \times \sum_{j,l=0}^{m-k,m'-k}\sqrt{\binom{m-k}{j}\binom{m'-k}{l}}\nonumber\\
&\times \ket{j,m-k-j} \bra{l,m'-k-l} \nonumber
\end{align}
The partial trace of $\rho_{AB}$ with respect to the subsystem $A$ yields the analytic expression for the linear entropy potential as
\begin{align}\nonumber
\label{sle}
L_E & = 1- N^4e^{-2|\alpha|^2}n!^2\sum_{m,m_1,m_2=0}^\infty \frac{|\alpha|^{2(m+m_2-2n+2k)}}{2^{m+m_2}m!m_2!} \nonumber\\ & \times \binom{m+m_2}{m_1}L_n^{m+k-n}(|\alpha|^2)L_n^{m_1+k-n}(|\alpha|^2) \nonumber\\ & \times L_n^{m_2+k-n}(|\alpha|^2)L_n^{m-m_1+m_2+k-n}(|\alpha|^2)
\end{align}
\begin{figure}[ht]
\centering
\includegraphics[scale=.4]{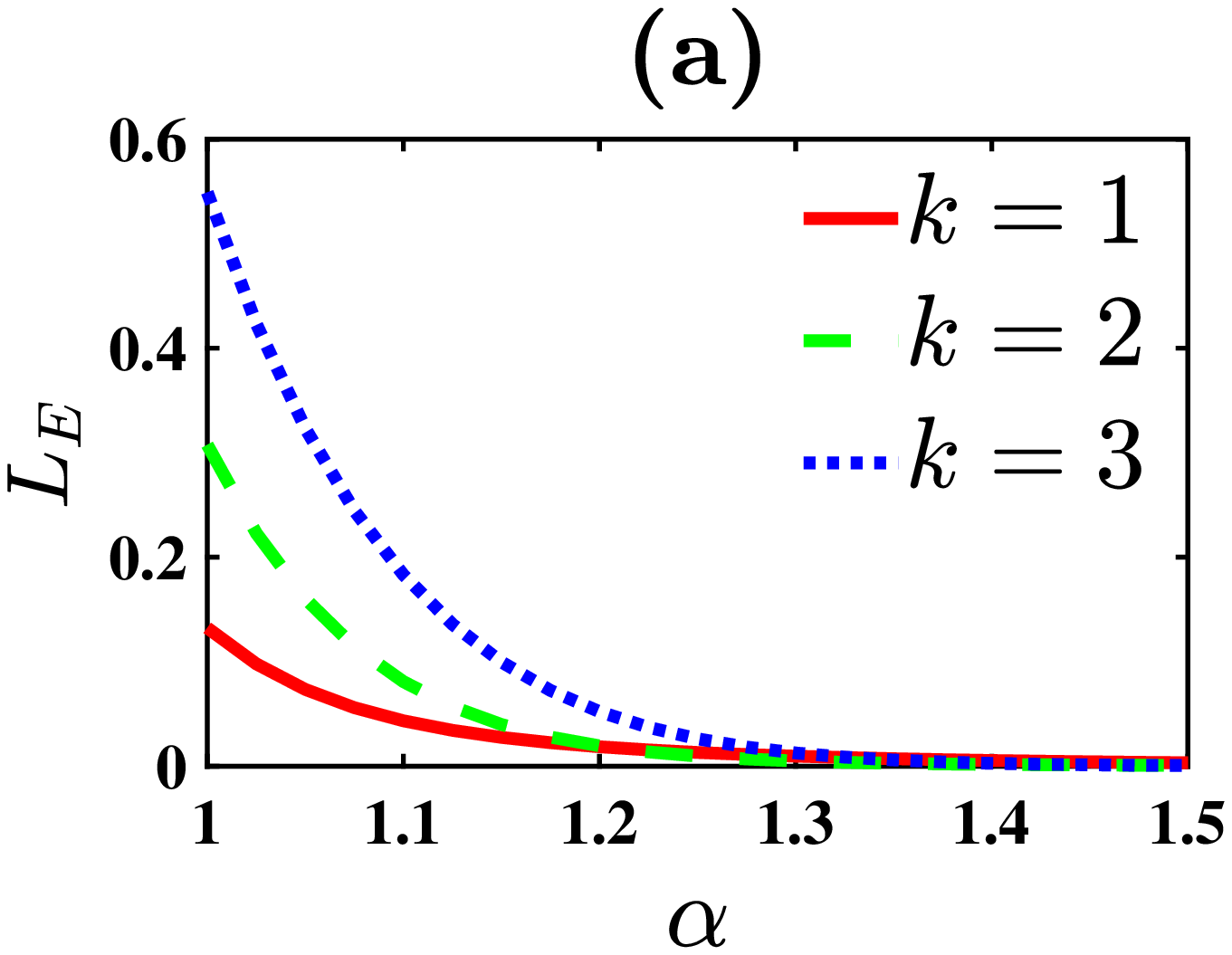}
\includegraphics[scale=.4]{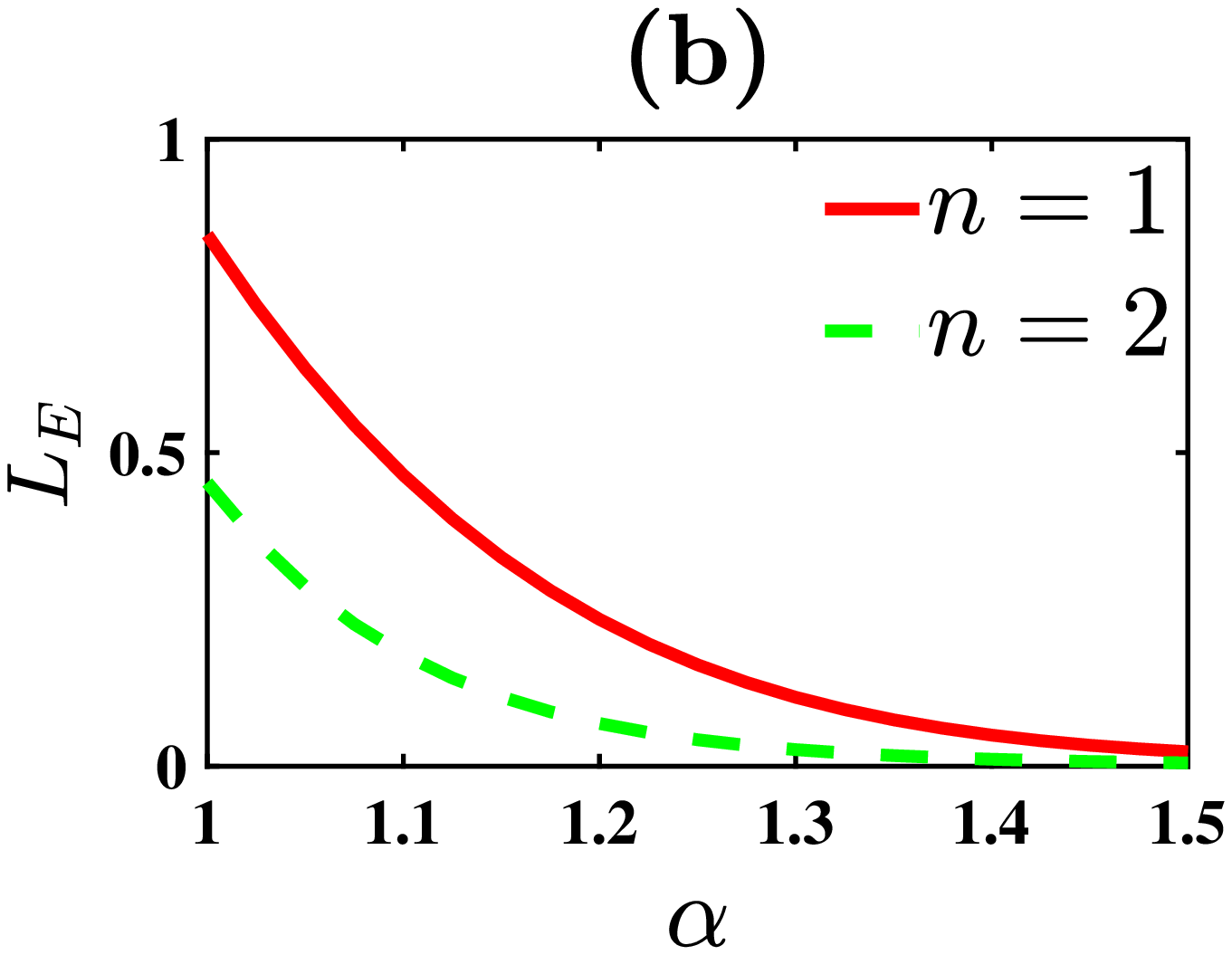}
\caption{(color online) Linear entropy $L_E$ for PSDFS as a function of displacement parameter $\alpha$ and for (a) different number of photon subtractions $k$ and $n=3$, (b) different values of Fock parameter $n$ and $k=1$.}
\label{figle}
\end{figure}

The variation in the amount of nonclassicality with respect to the state parameters is shown in Fig.~\ref{figle}. The nonclassicality of PSDFS is detected through the linear entropy potential. Here nonclassicality decreases with an increase in the displacement parameter as for $\alpha=0$ the state reduces to the Fock state, the most nonclassical state. The value of the displacement parameter is significant in controlling the effects of photon subtraction. In a certain range of $\alpha$, we can see that the amount of nonclassicality quantified by the linear entropy potential increases with the photon subtraction number for the states having the same initial Fock parameter. Interestingly, it is easy to observe from Fig.~\ref{figle}(b) that with an increase in the Fock parameter, the nonclassicality of PSDFS is found to be decreased. That means, photon
subtraction is an effective tool for enhancing nonclassicality in a certain range of $\alpha$ while higher values of the Fock parameter
are less beneficial in that range of the displacement parameter.

\subsection{Skew information-based measure}

In 2019, the skew information-based measure is proposed by Shunlong Luo et al. \cite{png49} in the context of Wigner Yanase skew information \cite{png50}. For a pure state $\rho$, it can be defined as
\begin{equation}
\label{isk}
N(\rho) = \frac{1}{2} + \langle a^\dagger a\rangle - \langle a^\dagger\rangle\langle a\rangle
\end{equation}
$N(\rho)$ represents the quantum coherence of $\rho$ with respect to the annihilation and creation operators, and is relatively easy to calculate. This measure is based on averages, which takes numerical value $\frac{1}{2}$ for classical coherent state and $n+\frac{1}{2}$ for mostly non-classical $n$ photon Fock
state $\ket{n}$. Thus any state $\rho$ with $N(\rho) > \frac{1}{2}$ is non-classical. However, this criterion is a one-sided condition as it fails for some Gaussian mixed states \cite{png49}.

In the case of PSDFS, $N(\rho)$ can be calculated using the general expectation of $a^{\dagger p}a^q$ as follows:
\begin{align}\nonumber
\label{eadpaq}
\langle a^{\dagger p}a^q \rangle &= \sum_{m,m'=0}^\infty C_m(n,\alpha,k)C_{m'}^*(n,\alpha,k)\bra{m'}a^{\dagger p}a^q\ket{m} \\& =
N^2e^{-|\alpha|^2}n!\alpha^{*p-q}|\alpha|^{2(k-n+q)}\sum_{m=0}^\infty \frac{|\alpha|^{2m}}{m!}\nonumber\\
&\times L_n^{m+q+k-n}(|\alpha|^2)L_n^{m+p+k-n}(|\alpha|^2)%\\ & =
\end{align}
where $L_n^\alpha(.)$ is the generalized Laguerre polynomial.
\begin{figure}[ht]
\centering
\includegraphics[scale=.4]{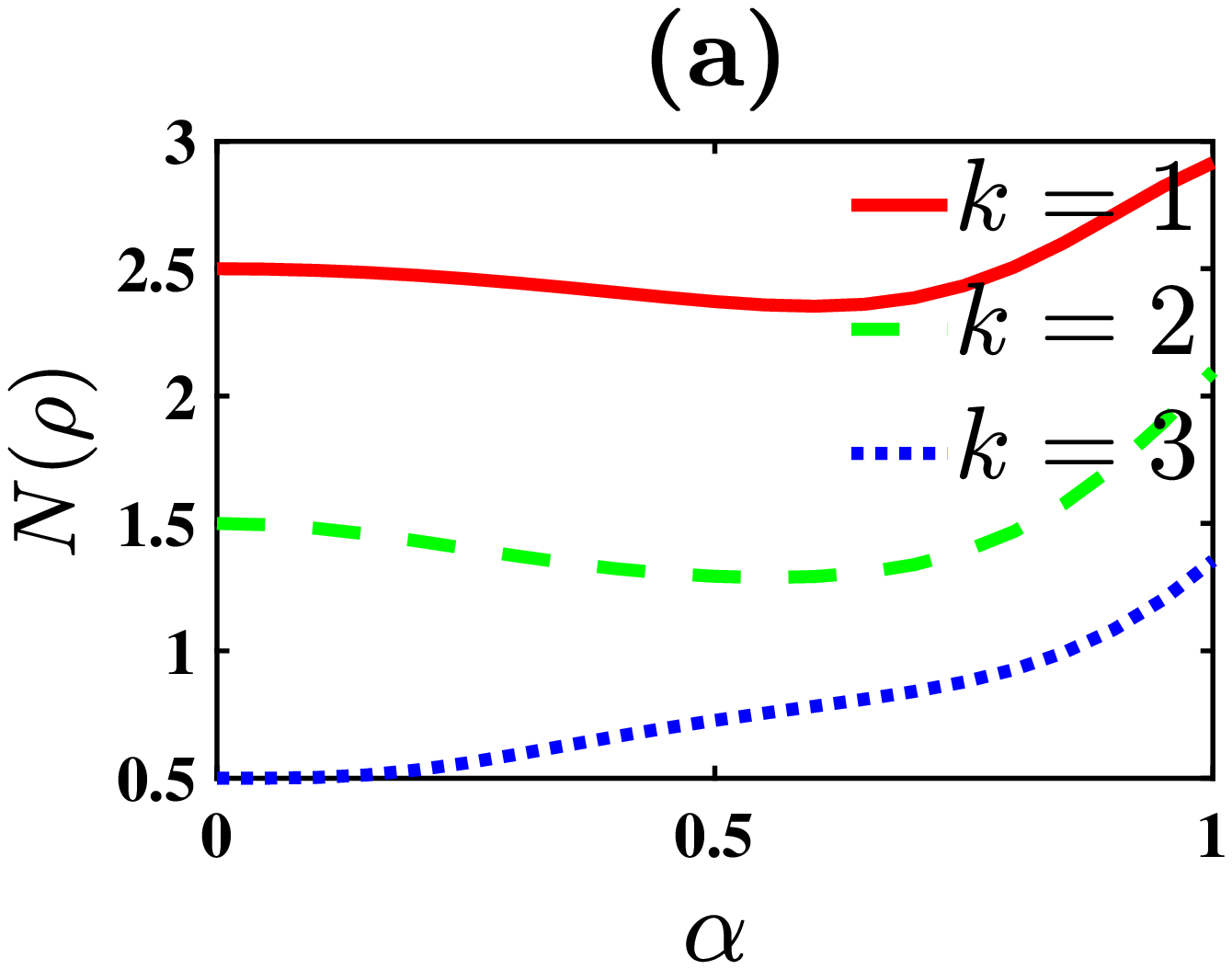}
\includegraphics[scale=.4]{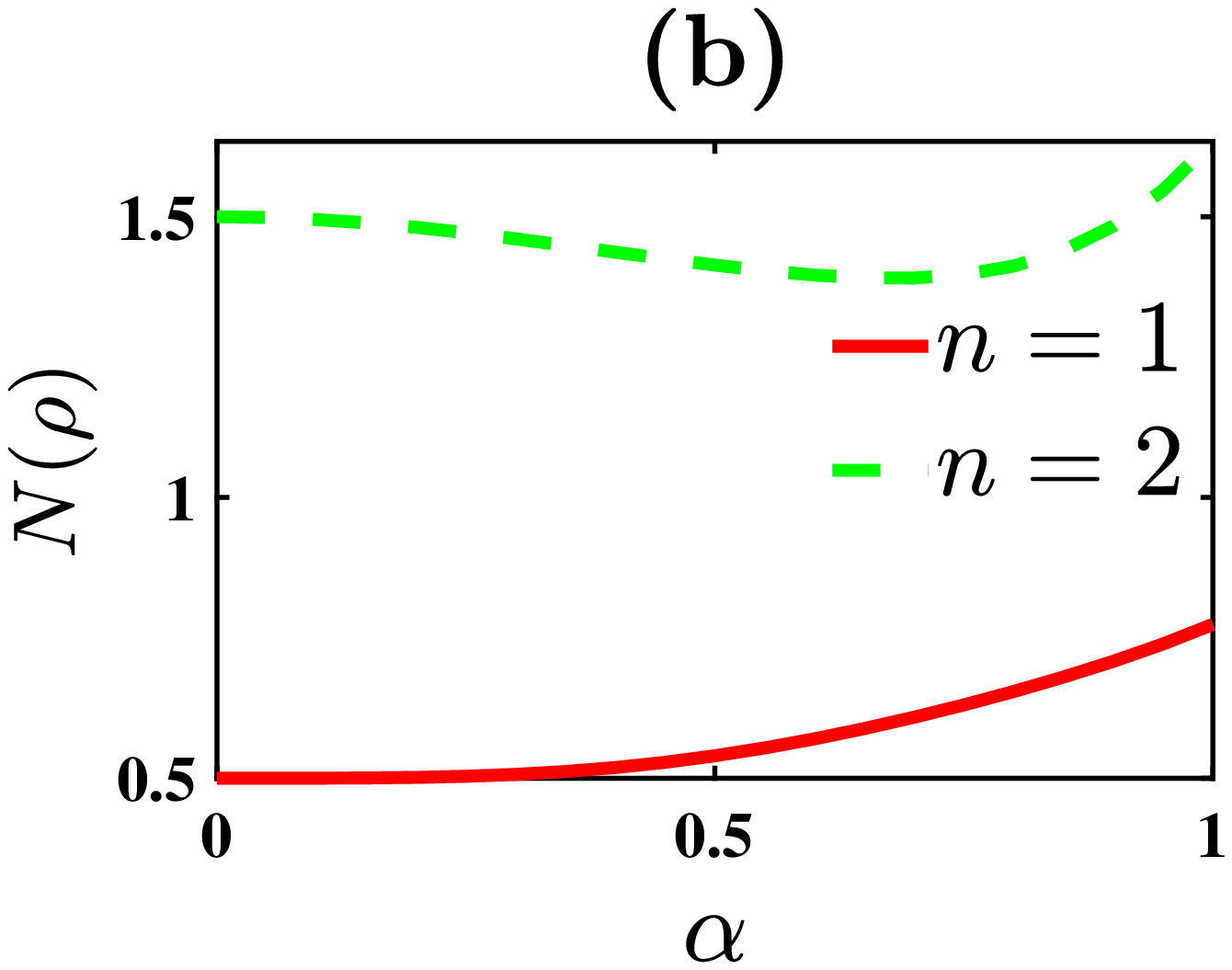}
\caption{(color online) A nonclassicality quantifier $N(\rho)$ with respect to the displacement parameter $\alpha$ and for (a) different photon subtraction $k$ and $n=3$ (b) different Fock parameter $n$ and $k=1$.}
\label{figsibm}
\end{figure}

Figure~\ref{figsibm}(a) shows that the nonclassicality expressed by the skew information-based measure cannot be enhanced by increasing the displacement parameter. It can also be observed that with an increase in photon subtraction, nonclassicality remains almost the same up to a certain value of displacement parameter and decreases thereafter. Moreover, Fig.~\ref{figsibm}(b) shows that the nonclassicality increases with an increase in the Fock parameter, which is the opposite as noted earlier in the case of linear entropy potential (see Fig.~\ref{figle}(b)), indicating that the validity of any such ``amount of nonclassicality" based ordering is restricted to the specific measure involved.

\subsection{Wigner logarithmic negativity}

We have seen in Fig.~\ref{figwf} that the Wigner function diagnoses the non-classical as well as the quantum non-Gaussian nature of the PSDFS. This inspired us to quantify the amount of nonclassicality and quantum non-Gaussianity using the negative volume of the Wigner function. A simple measure of nonclassicality, named Wigner logarithmic negativity is introduced as
\cite{al}
\begin{equation}
\label{iwln}
W = \log_2\left(\int{d^2\gamma}|W(\gamma,\gamma^*)| \right)
\end{equation}
where the integration is performed over the complete region $\mathcal{R}^{2n}$ of phase-space. More interestingly, in the resource theory of quantum information, it is
noted that $W$ also estimates the amount of quantum non-Gaussianity present in the PSDFS as the negative values of the Wigner function also witness the quantum non-Gaussianity of the field state.

The integration in \eqref{iwln} is executed numerically using the Wigner function \eqref{swf} to study the effect of different parameters on
the Wigner logarithmic negativity of PSDFS.
\begin{figure}[ht]
\centering
\includegraphics[scale=.4]{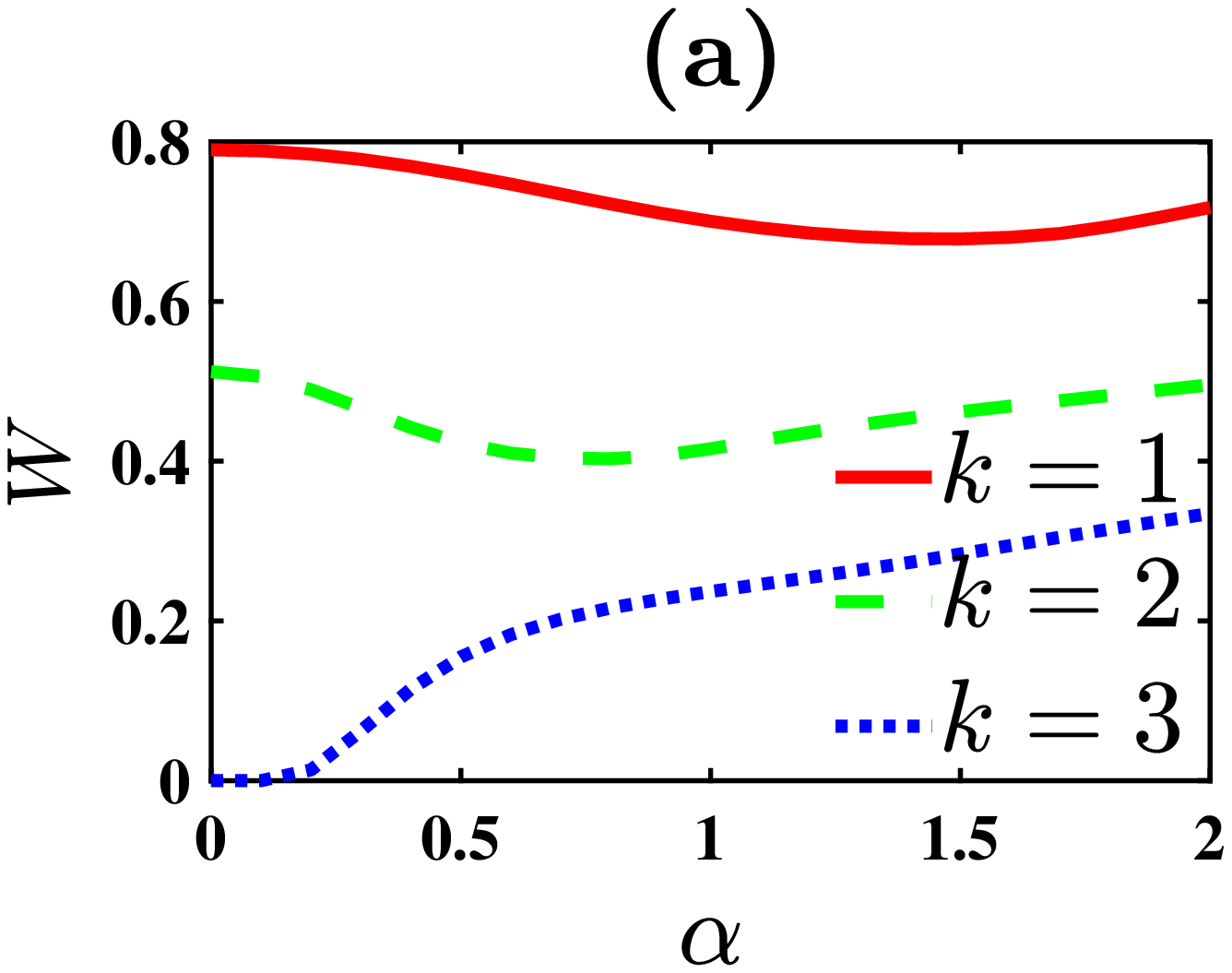}
\includegraphics[scale=.4]{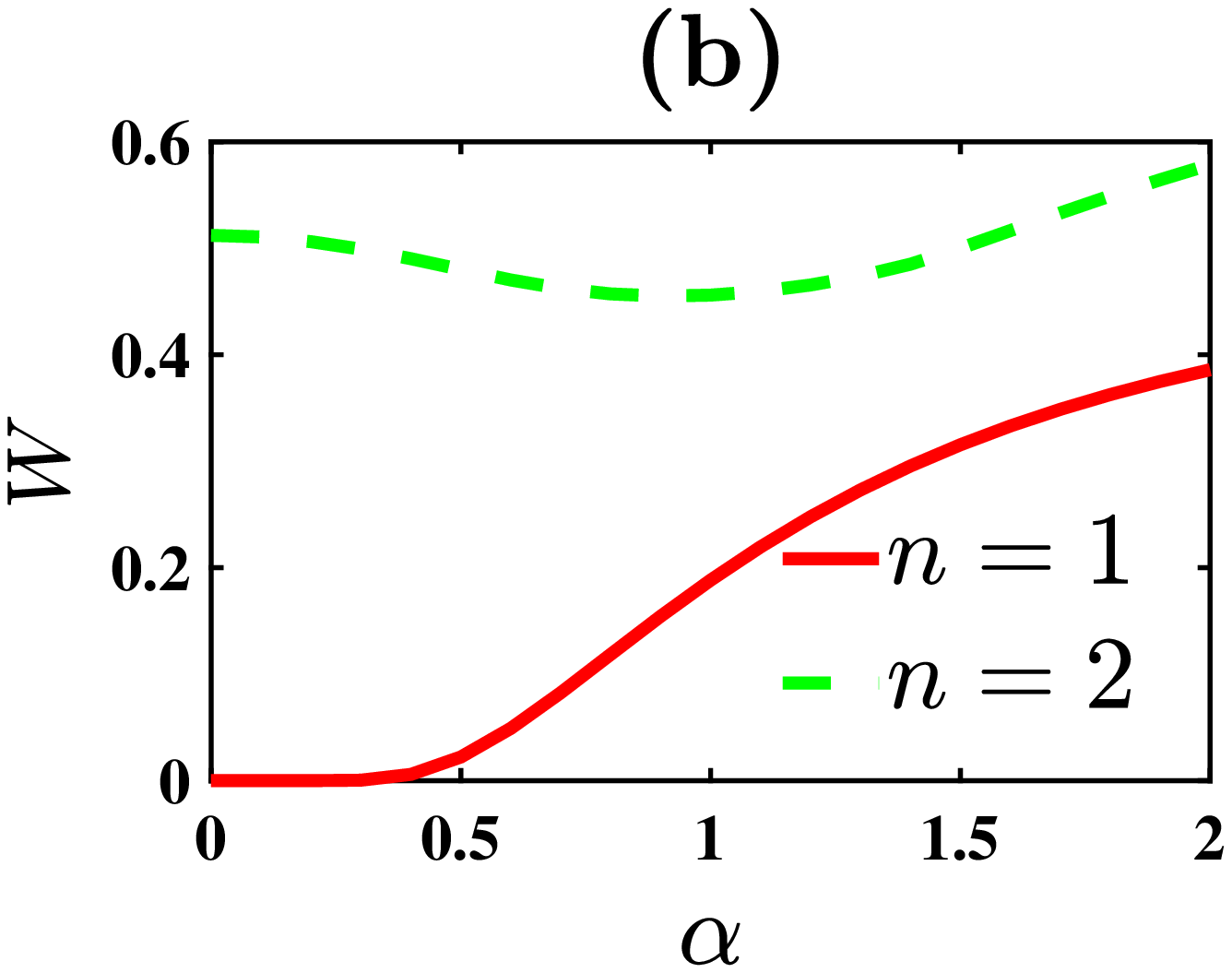}
\caption{(color online) Changes of Wigner logarithmic negativity $W$ with respect to the displacement parameter $\alpha$ for (a) different values of $k$ and $n=3$ (b) different values of $n$ and $k=1$.}
\label{figwln}
\end{figure}

In the case of increasing photon subtraction number $k$ from 1 to 3, the Wigner logarithmic negativity decreases.  Specifically, any increase in the displacement
parameter only rises nonclassicality and then quantum non-Gaussianity. Among the photon subtraction and Fock parameter, the second one is more effective in enhancing the nonclassicality and quantum non-Gaussianity of PSDFS when a single photon is subtracted.

\subsection{Relative entropy measure of quantum non-Gaussianity}

Different measures to inspect the quantum non-Gaussianity of a quantum state have been reported in literature \cite{al, png21, png52, png53, png54}. In the previous section, we have observed that in the case of PSDFS, the variation in the quantum non-Gaussianity with respect to different state parameters is similar to that of nonclassicality as revealed by the Wigner logarithmic negativity.  This is justified because states with negative Wigner function are a subset of quantum non-Gaussian states \cite{al}. Hence we further investigate the quantum non-Gaussianity of the considered state using the relative entropy of quantum non-Gaussianity.

This important measure can be defined with the set of all Gaussian states as follows \cite{genoni2008quantifying}%\cite{png}
\begin{equation}
\label{imng}
\delta[\rho]=S(\rho||\tau_G)
\end{equation}
where the relative entropy is $S(\rho||\Lambda) =\Tr[\rho(\log(\rho)-\log(\Lambda))]$, and the reference mixed Gaussian state $\tau_G$ is selected in such a way that the first and second-order moments are same as that of $\rho$. Further, in the case of the Gaussian state, $S(\rho=\ket\psi\bra\psi)$ is zero, and hence $\delta[\ket\psi]=S(\tau_G)$, which is the von Neumann entropy of the reference state.

A Gaussian state is fully characterized by its two first-order moments and hence by its covariance matrix which can be written as \cite{al}
\begin{equation}
\label{matrix}
\sigma = \left[
\begin{matrix}
\sigma_{pp} & \sigma_{qp} \\  \sigma_{qp} &\sigma_{pp}
\end{matrix}
\right]
\end{equation}
where $\sigma_{pq}=\langle{pq+qp}\rangle -2\langle{p}\rangle\langle q\rangle$ for position $q=\frac{a+a^\dagger}{\sqrt{2}}$ and momentum $p=\frac{a-a^\dagger}{i\sqrt{2}}$. Using \eqref{eadpaq}, all the elements of the covariance matrix $\sigma$ for PSDFS can be obtained and thus the relative entropy measure of quantum non-Gaussianity reduces to \cite{png}
\begin{equation}
\label{rmng}
\delta[\ket\psi]=S(\tau_G) = h(\det(\sqrt{\sigma}))
\end{equation}
with $h(x)=\frac{x+1}{2}\log_2\left(\frac{x+1}{2}\right)-\frac{x-1}{2}\log_2\left(\frac{x-1}{2}\right)$.
\begin{figure}[ht]
\centering
\includegraphics[scale=.4]{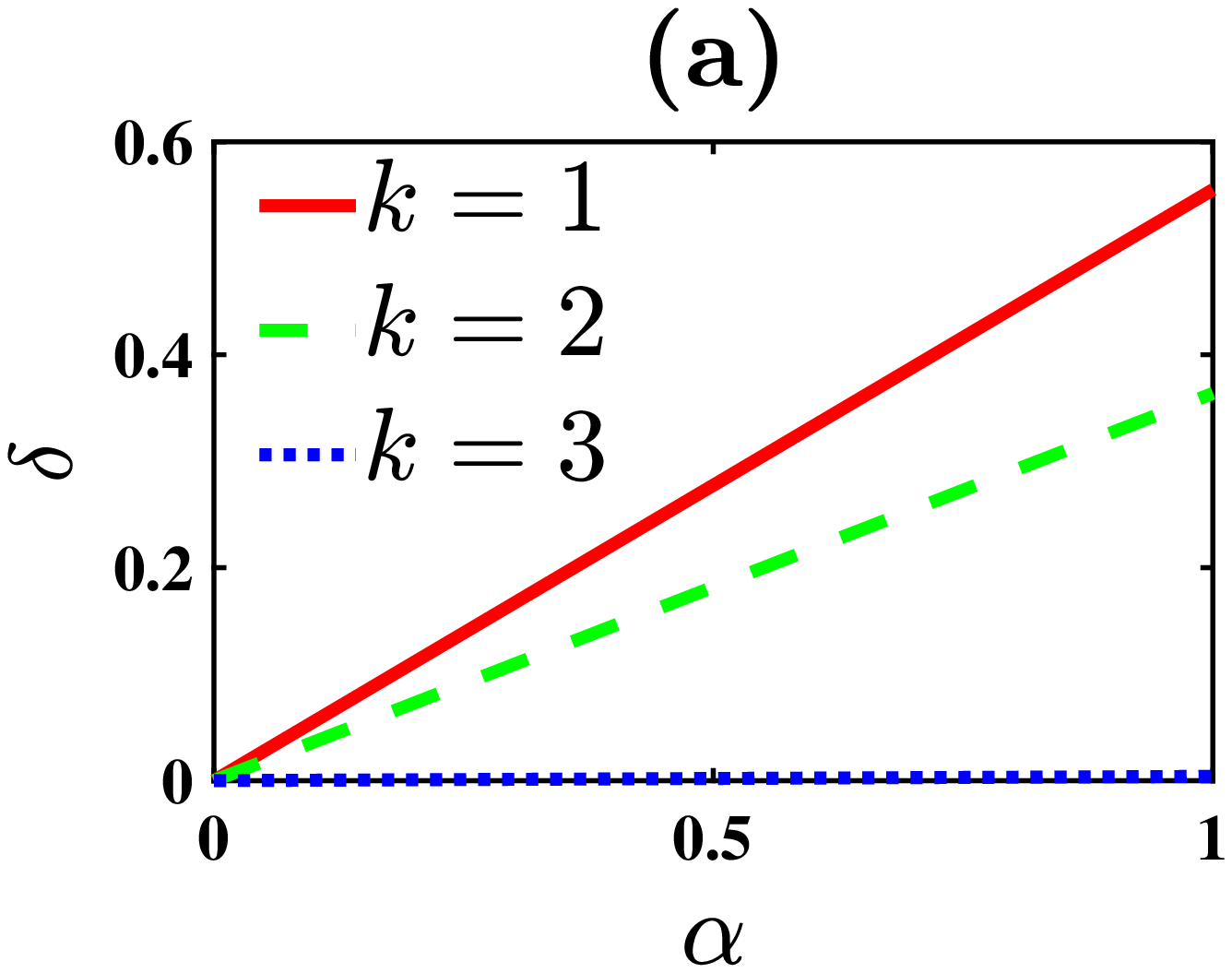}
\includegraphics[scale=.4]{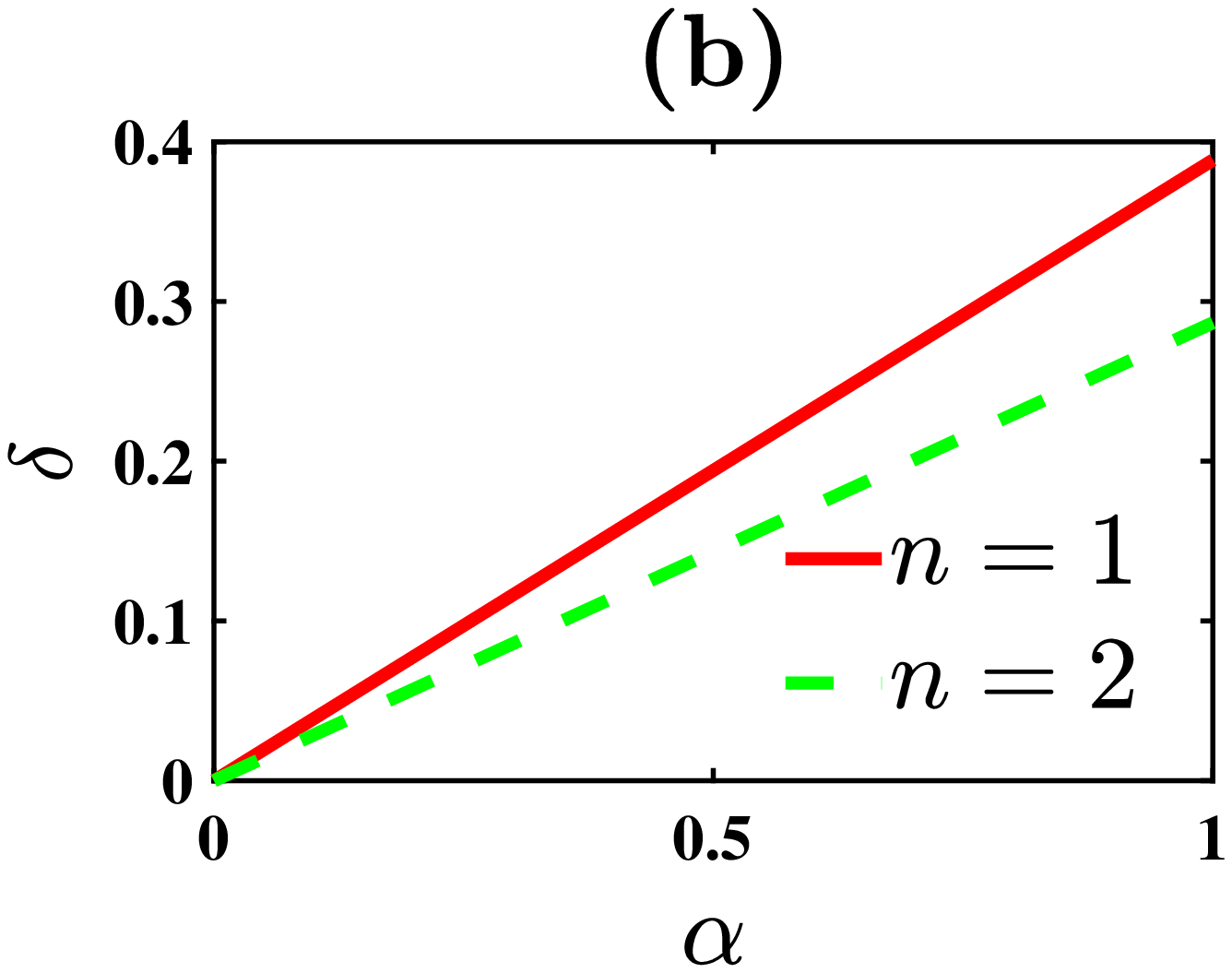}
\caption{(color online) quantum non-Gaussianity as a function of $\alpha$ for (a) variation in $k$ and $n=3$ (b) different Fock parameter $n$ values and $k=1$.}
\label{figmng}
\end{figure}

It can be seen from Fig.~\ref{figmng}{\color{blue}(a)} that the plots are increasing with $\alpha$ for $k=1,\,2$, and become zero for $k=3$. Also, the plots are gradually approaching the horizontal axis. Consequently, we can conclude that subtraction of photons reduces the relative entropy to zero or equivalently application of quantum non-Gaussianity inducing annihilation operator thrice increases quantum non-Gaussianity and nonclassicality of PSDFD for relatively small values of $\alpha$. Also Fig.~\ref{figmng} {\color{blue}(b)} shows that relative entropy increase with increasing Fock state parameter $n$.

\section{Wigner function of PSDFS evolving under photon loss channel}
\label{loss}

The interaction between a quantum system and its environment instigates quantum to classical transition. So, the observed non-classical and quantum non-Gaussian features of the PSDFS are expected to deteriorate due to its evolution under the lossy channel. In particular, the temporal evolution of a quantum state $\rho$ over the lossy channel can be studied using the LGKS master equation \cite{png57} given by
\begin{equation}
\label{lgksme}
\frac{\partial\rho}{\partial t} = \kappa(2a\rho a^\dagger -a^\dagger a\rho - \rho a^\dagger a)
\end{equation}
where $\kappa$ is the rate of decay. Analogously, the time evolution of the Wigner function at time $t$ in terms of the initial Wigner function of the state which is evolving under the lossy channel \cite{png58} is defined as
\begin{equation}
\label{dwlc}
W(\zeta,t) = \frac{2}{T}\int\frac{\partial^2\gamma}{\pi}\exp[-\frac{2}{T}|\zeta-\gamma e^{-\kappa t}|^2]W(\gamma,\gamma^*,0)
\end{equation}
with $T = 1- \exp(-2\kappa t)$ and $W(\gamma,\gamma^*,0)$ is the Wigner function at initial time $t=0$ which is calculated in \eqref{swf}. The time evolution of Wigner function \eqref{dwlc} models dissipation due to interaction with a vacuum reservoir as well as inefficient detectors with efficiency $\eta=1-T$. The detailed calculation of $W(\zeta,t)$ is given in Appendix \ref{awflc}.
\begin{figure*}[htb]
\centering
\includegraphics[scale=1]{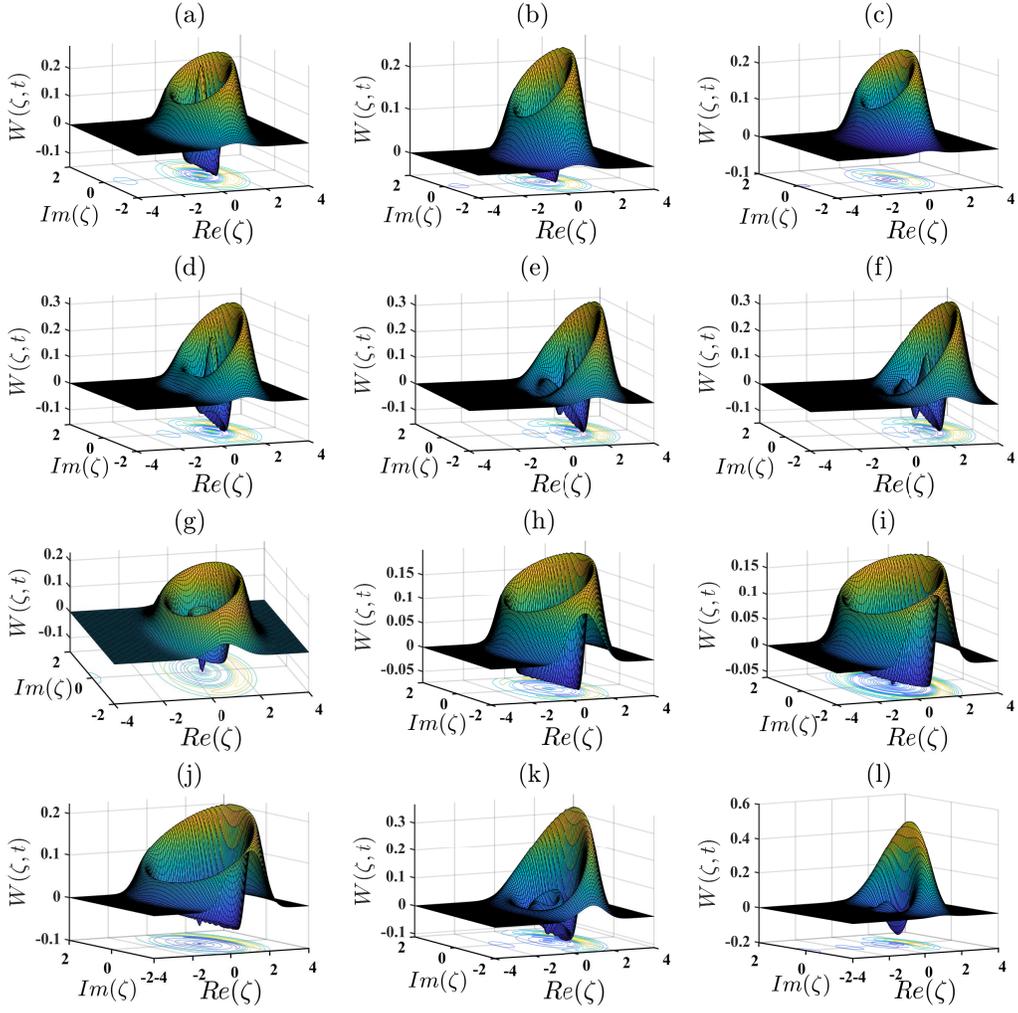}
\caption{(Color online) The dynamics of Wigner function evolving under photon loss channel for $k=1$, $n=3$, $\alpha=0.5$ and with different values of the rescaled time (a) $\kappa t = 0.1$, (b) $\kappa t = 0.3$, (c) $\kappa t = 0.5$; $k=1$, $n=3$, $\kappa t=0.1$ and with different values of displacement parameter (d) $\alpha=1$, (e) $\alpha=1.5$, (f) $\alpha=2$; $k=1$, $\kappa t=0.1$, $\alpha=0.5$ and with different values of Fock state parameter (g) $n=4$, (h) $n=5$, (i) $n=6$; $n=3$, $\kappa t=0.1$, $\alpha=0.5$ and with different values of photon subtraction parameter (j) $k=2$, (k) $k=4$, (l) $k=6$, respectively.}
\label{figwflc}
\end{figure*}

A compact expression of the Wigner function evolving under the photon loss channel can be obtained from \eqref{dwlc} by using the description of the Wigner function in an ideal situation.
One can easily notice that both the quantum non-Gaussianity and nonclassicality of PSDFS cannot increase due to its temporal evolution over a photon loss channel. This is clearly illustrated by Fig.~\ref{figwflc}(a)-(c) that the negative region is shrinking with increasing values of rescaled time $\kappa t$ for a single photon subtracted DFS with $n=3$ and displacement parameter $\alpha = 0.5$. That means increasing $\kappa t$ diminishes the quantum features of the PSDFS. Moreover, it is clear from Fig.~\ref{figwflc}(d)-(f) that the displacement parameter $\alpha$ is not a dominating factor for the nonclassicality and quantum non-Gaussianity of PSDFS under the effect of photon loss channel. Interestingly,  the Wigner function for PSDFS evolving under photon loss channel behaves unlikely as the variation of the Wigner function in the absence of noise with the Fock parameter. Both the nonclassicality and quantum non-Gaussianity decrease as $n$ increases.

\section{Conclusion}
\label{conclusion}

The description as well as the quantification of nonclassicality and quantum non-Gaussianity of quantum states are problems of interest in their merit. These problems are addressed here by considering PSDFS as a test case. Specifically, PSDFS is chosen as it can be reduced to various quantum states having important applications in different limits. In this work, we have used linear entropy potential, skew information-based criteria, and Wigner logarithmic negativity as measures of nonclassicality to compare the
amount of nonclasicality present in a set of different quantum non-Gaussian pure states. The nonclassicality and quantum non-Gaussianity present in the PSDFS are quantified, which shows that both photon subtraction and Fock parameters enhance these quantum features whereas the latter is a more effective tool at small displacement
parameters. In contrast, the displacement parameter is observed to reduce the quantum features. In view of Hudson's theorem and
resource theory of quantum non-Gaussianity based on Wigner negativity, the Gaussian operations are free, and thus, displacement operation and photon loss channels are not expected to enhance quantum non-Gaussianity and/or Wigner negativity. Although, in the case of pure PSDFS, the Wigner negativity captures all the quantum non-Gaussianity but cannot predict conclusively the feasibility of quantum non-Gaussianty in PSDFS measured through inefficient detectors (or evolved through lossy channels)
with the positive Wigner function. Similarly, skew information-based and linear entropy potential measures show that the Wigner function succeeds in detecting all the nonclassicality of the pure PSDFS.

{ In this context, it can be noted that the non-Gaussianity of PSDFS for non-zero values of $n,\,\alpha$ and $k$ is detected by the non-Gaussianity identifying Wigner function and it decreases with the increasing values of $n,\,\alpha$ and $k$. The amount of non-Gaussianity is quantified by relative entropy $\delta$ which increases with $\alpha$ but decreases with $n$ and $k$. However, nonclassicality recognized by skew information-based measure is decreasing with $k$ and increasing with $n$ and $\alpha$. The amount of nonclassicality present in PSDFS can be estimated by using different measures like linear entropy and Wigner logarithmic negativity but they are not consistent with respect to different parameters. The amount of nonclassicality assessed by the 
linear entropy is increasing with $k$ and decreasing with $n$ and $\alpha$ while Wigner logarithmic negativity behaves just opposite that means it decreases with $k$ and increases with $n$ and $\alpha$.}

Following the recent progress in quantum state engineering and quantum information processing, we conclude this article with a belief that PSDFSs will soon be constructed in laboratories and will be used for performing quantum optical, metrological, and computing tasks.

\begin{widetext}
\begin{appendix}
\section{ Detailed Simplification of state PSDFS}
\label{appdspsdfs}
We have,
\begin{align*}
\ket{\psi} & = Na^kD(\alpha)\ket{n} =Ne^{-\frac{|\alpha|^2}{2}}a^ke^{a^\dagger\alpha}e^{-a\alpha^*}\ket{n} \\& =
Ne^{-\frac{|\alpha|^2}{2}}a^ke^{a^\dagger\alpha}\sum_{p=0}^n\frac{(-\alpha^*)^p}{p!}a^p\ket{n} \\& =
Ne^{-\frac{|\alpha|^2}{2}}a^ke^{a^\dagger\alpha}\sum_{p=0}^n\frac{(-\alpha^*)^p}{p!}\sqrt{\frac{n!}{(n-p)!}}\ket{n-p} \\&
=Ne^{-\frac{|\alpha|^2}{2}}a^k\sum_{q=0}^\infty\frac{\alpha^q}{q!}a^{\dagger q}\sum_{p=0}^n\frac{(-\alpha^*)^p}{p!}\sqrt{\frac{n!}{(n-p)!}}\ket{n-p} \\& =
Ne^{-\frac{|\alpha|^2}{2}}\sum_{q=0}^\infty\frac{\alpha^q}{q!}\sum_{p=0}^n\frac{(-\alpha^*)^p}{p!}\sqrt{\frac{n!}{(n-p)!}}a^ka^{\dagger q}\ket{n-p} \\& =
Ne^{-\frac{|\alpha|^2}{2}}\sum_{q=0}^\infty\frac{\alpha^q}{q!}\sum_{p=0}^n\frac{(-\alpha^*)^p}{p!}\sqrt{\frac{n!}{(n-p)!}}\sqrt{\frac{(n-p+q)!}{(n-p)!}} a^k\ket{n-p+q} \\& =
Ne^{-\frac{|\alpha|^2}{2}}\sum_{q=0}^\infty\frac{\alpha^q}{q!}\sum_{p=0}^n\frac{(-\alpha^*)^p}{p!}\frac{\sqrt{n!(n-p+q)!}}{(n-p)!}\sqrt{\frac{(n-p+q)!}{(n-p+q-k)!}} \ket{n-p+q-k} \\& =
Ne^{-\frac{|\alpha|^2}{2}}\sum_{q=0}^\infty\frac{\alpha^q}{q!}\sum_{p=0}^{\min(n,n-k+q)}\frac{(-\alpha^*)^p}{p!}\frac{{(n-p+q)!}}{(n-p)!}\sqrt{\frac{n!}{(n-p+q-k)!}} \ket{n-p+q-k} \\& =
Ne^{-\frac{|\alpha|^2}{2}}\sum_{m=k}^\infty\sum_{p=0}^n\frac{\alpha^{m-n+p}}{(m-n+p)!}\frac{(-\alpha^*)^p}{p!}\frac{{m!}}{(n-p)!}\sqrt{\frac{n!}{(m-k)!}} \ket{m-k} \\& =
Ne^{-\frac{|\alpha|^2}{2}}\sum_{m=k}^\infty\alpha^{m-n}\sqrt{\frac{n!}{(m-k)!}}L_n^{m-n}(|\alpha|^2)\ket{m-k} \\ & =
\sum_{m=k}^\infty C_m(n,\alpha,k)\ket{m-k}
\end{align*}
where, $C_m(n,\alpha,k) = Ne^{-\frac{|\alpha|^2}{2}}\alpha^{m-n}\sqrt{\frac{n!}{(m-k)!}}L_n^{m-n}(|\alpha|^2)$.
\section{Wigner function of PSDFS}
\label{awf}
We have $\ket\psi = Na^kD(\alpha)\ket{n}$. Using this
\begin{align*}
& C(\lambda)\\ & = \text{Tr}(\rho D(\lambda))\\ & = N^2\bra{n}D^\dagger(\alpha)a^{\dagger k}D(\lambda)a^kD(\alpha)\ket{n} \\ & =
N^2\bra{n}(a^\dagger+\alpha^*)^kD^\dagger(\alpha)D(\lambda)D(\alpha)(a+\alpha)^k\ket{n} \\& =
N^2\exp(\lambda\alpha^*-\lambda^*\alpha)\sum_{p,q=0}^k\binom{k}{p}\binom{k}{q}\alpha^{*k-p}\alpha^{k-q}\bra{n}a^{\dagger p}D(\lambda) a^q\ket{n} \\& =
N^2\exp(\lambda\alpha^*-\lambda^*\alpha-\frac{|\lambda|^2}{2})\sum_{p,q=0}^k\binom{k}{p}\binom{k}{q}\alpha^{*k-p}\alpha^{k-q}n!\sum_{r=0}^{n-p}\frac{\lambda^r(-\lambda^*)^{p-q+r}}{r!(p-q+r)!(n-p-r)!}
\end{align*}
Now the Wigner function is given by
\begin{align*}
& W(\gamma,\gamma^*)\\
& = \frac{1}{\pi^2}\int{d^2\lambda}\,C(\lambda)e^{\gamma\lambda^*-\gamma^*\lambda} \\& =
N^2\sum_{p,q=0}^k\binom{k}{p}\binom{k}{q}\alpha^{*k-p}\alpha^{k-q} n!\sum_{r=0}^{n-p}\frac{1}{r!(p-q+r)!(n-p-r)!}\\ & \times
\frac{1}{\pi^2}\int{d^2\lambda}\,\lambda^r(-\lambda^*)^{p-q+r}\exp(\lambda\alpha^*-\lambda^*\alpha-\frac{|\lambda|^2}{2}+\gamma\lambda^*-\gamma^*\lambda) \\& =
N^2\sum_{p,q=0}^k\binom{k}{p}\binom{k}{q}\alpha^{*k-p}\alpha^{k-q} n!\sum_{r=0}^{n-p}\frac{1}{r!(p-q+r)!(n-p-r)!}\\ & \times \frac{1}{\pi^2}\int{d^2\lambda}\,\lambda^r(-\lambda^*)^{p-q+r}\exp(\lambda\eta^*-\lambda^*\eta-\frac{|\lambda|^2}{2})~~~~~\mbox{substituing }\eta=\alpha-\gamma\\ & =
N^2\sum_{p,q=0}^k\binom{k}{p}\binom{k}{q}\alpha^{*k-p}\alpha^{k-q} n!\sum_{r=0}^{n-p}\frac{1}{r!(p-q+r)!(n-p-r)!}\\ & \times \frac{\partial^r}{\partial\eta^{*r}}\frac{\partial^{p-q+r}}{\partial\eta^{p-q+r}}\frac{1}{\pi^2}\int{d^2\lambda}\,\exp(\lambda\eta^*-\lambda^*\eta-\frac{|\lambda|^2}{2})\\& =
N^2\sum_{p,q=0}^k\binom{k}{p}\binom{k}{q}\alpha^{*k-p}\alpha^{k-q} n!\sum_{r=0}^{n-p}\frac{1}{r!(p-q+r)!(n-p-r)!} \frac{\partial^r}{\partial\eta^{*r}}\frac{\partial^{p-q+r}}{\partial\eta^{p-q+r}}\frac{2}{\pi}\exp(-2|\eta|^2)\\& =
\frac{2N^2}{\pi}\sum_{p,q=0}^k\binom{k}{p}\binom{k}{q}\alpha^{*k-p}\alpha^{k-q} n!\sum_{r=0}^{n-p}\frac{(-2)^{p-q+r}\eta^{*p-q}}{r!(p-q+r)!(n-p-r)!} \\ & \times \sum_{s=0}^r\binom{r}{s} \frac{(p-q+r)!}{(p-q+r-s)!}|\eta|^{2(r-s)}(-2)^{r-s}\exp(-2|\eta|^2)\\& =
\frac{2N^2\exp(-2|\eta|^2)}{\pi}\sum_{p,q=0}^k\binom{k}{p}\binom{k}{q}\alpha^{*k-p}\alpha^{k-q} n!\\&\times\sum_{r=0}^{n-p}\frac{(-2)^{p-q+r}\eta^{*p-q}}{r!(p-q)!(n-p-r)!}~_1F_1(-r;p-q+1;2|\eta|^2)
\end{align*}
\section{Wigner function evolving under photon loss channel}
\label{awflc}
\begin{align*}
& W(\zeta,t)\\ & = \frac{2}{T}\int\frac{d^2\gamma}{\pi}\exp[-\frac{2}{T}|\zeta-\gamma e^{-\kappa t}|^2] W(\gamma, \gamma^*, 0)\\& =
\frac{2}{T}\frac{2N^2}{\pi}\sum_{p,q=0}^k\binom{k}{p}\binom{k}{q}\alpha^{*k-p}\alpha^{k-q} n!\sum_{r=0}^{n-p}\frac{(-2)^{p-q+r}}{r!(n-p-r)!}\sum_{s=0}^r\binom{r}{s} \frac{(-2)^{r-s}}{(p-q+r-s)!} \\ & \times \int\frac{d^2\gamma}{\pi}\exp[-\frac{2}{T}|\zeta-\gamma e^{-\kappa t}|^2]\eta^{*p-q}|\eta|^{2(r-s)}\exp(-2|\eta|^2) \\& =
\frac{2}{T}\frac{2N^2}{\pi}\sum_{p,q=0}^k\binom{k}{p}\binom{k}{q}\alpha^{*k-p}\alpha^{k-q} n!\sum_{r=0}^{n-p}\frac{(-2)^{p-q+r}}{r!(n-p-r)!}\sum_{s=0}^r\binom{r}{s} \frac{(-2)^{r-s}}{(p-q+r-s)!} \\ & \times \int\frac{d^2\eta}{\pi}\exp[-\frac{2}{T}\left[|\beta|^2e^{2\kappa t}+\beta\eta^*+\beta^*\eta +|\eta|^2(T+e^{-2\kappa t})\right]]\eta^{*p-q+r-s}\eta^{r-s}~\mbox{assuming $\zeta-\alpha e^{-\kappa t}=\beta e^{\kappa t}$} \\&=
\frac{2}{T}\frac{2N^2}{\pi}\sum_{p,q=0}^k\binom{k}{p}\binom{k}{q}\alpha^{*k-p}\alpha^{k-q} n!\sum_{r=0}^{n-p}\frac{(-2)^{p-q+r}}{r!(n-p-r)!}\sum_{s=0}^r\binom{r}{s} \frac{(-2)^{r-s}}{(p-q+r-s)!}\exp(-\frac{2}{T}|\beta|^2e^{2\kappa t}) \\ & \times\left(-\frac{T}{2}\right)^{p-q+2(r-s)}\frac{\partial^{p-q+r-s}}{\partial\beta^{p-q+r-s}}\frac{\partial^{r-s}}{\partial\beta^{*r-s}}\int\frac{d^2\eta}{\pi}
\exp[-\frac{2}{T}(\beta\eta^*+\beta^*\eta +|\eta|^2)]~~~\mbox{using $T=1-e^{-2\kappa t}$} \\&= \frac{2N^2}{\pi}
n!\exp[\frac{2}{T}|\beta|^2(1-e^{2\kappa t})]\sum_{p,q=0}^k\binom{k}{p}\binom{k}{q}\alpha^{*k-p}\alpha^{k-q}\left(2\beta^*\right)^{p-q}\sum_{r=0}^{n-p}\frac{(-2)^{r}}{(n-p-r)!} \\ & \times \sum_{s=0}^r \frac{(-2)^{r-s}}{s!}\sum_{u=0}^{p-q+r-s}\frac{|\beta|^{2(r-s-u)}}{u!(p-q+r-s-u)!(r-s-u)!}\left(\frac{2}{T}\right)^{-u} % \\ &=
\end{align*}
\end{appendix}
\end{widetext}
\begin{center}
\textbf{ACKNOWLEDGEMENT}
\end{center}

Deepak's work is supported by the Council of Scientific and Industrial Research (CSIR), Govt. of India (Award no. 09/1256(0006)/2019-EMR-1). A. C. acknowledges SERB, DST for the support provided through the project number SUR/2022/000899.  
\bibliographystyle{unsrt}
\bibliography{ref}

\end{document}